\documentclass[a4paper,10pt]{article}
\usepackage[utf8]{inputenc}
\usepackage{graphicx}
\usepackage{amssymb}
\usepackage{amsmath} 
\usepackage{bm}

\renewcommand{\vec}[1]{{\bm{#1}}}
\newcommand{\fr}[2]{{\displaystyle \frac{#1}{#2}}}
\newcommand{\sfr}[2]{{{#1}/{#2}}}
\newcommand{\diff}[2]{{\fr{d{#1}}{d{#2}}}}
\newcommand{\pdiff}[2]{{\fr{\partial{#1}}{\partial{#2}}}}

\newcommand{\spdiff}[2]{{\sfr{\partial{#1}}{\partial{#2}}}}

\newcommand{\rotor}{\mathop{\rm rot}\nolimits}
\newcommand{\Rm}{{\rm R}_{\rm m}}

\title{Features of the Accretion in the EX Hydrae System: Results of Numerical Simulation}
\author{P. B. Isakova$^{1}$\thanks{isakovapb@inasan.ru},
        A. G. Zhilkin$^{1}$,
        D. V. Bisikalo$^{1}$,\\
        A. N. Semena$^{2}$, and
        $\fbox{M. G. Revnivtsev}^{2}$\\
        \textit{\small $^{1}$ Institute of Astronomy RAS, Moscow, Russia}\\
        \textit{\small $^{2}$ Space Research Institute, Moscow, Russia}}
\date{}
\begin{document}

\maketitle

\begin{abstract}

A two dimensional numerical model in the axisymmetric approximation that describes the flow structure in the magnetosphere of the white dwarf in the EX Hya system has been developed. Results of simulations show that the accretion in EX Hya proceeds via accretion columns, that are not closed and
have curtain-like shapes. The thickness of the accretion curtains depends only weakly on the thickness
of the accretion disk. This thickness developed in the simulations does not agree with observations. It is concluded that the main reason for the formation of thick accretion curtains in the used model is the assumption that the magnetic field penetrates fully into the plasma of the disk. An analysis based on simple estimates shows that a diamagnetic disk that fully or partially shields the magnetic field of the star may be a more attractive explanation for the observed features of the accretion in EX Hya.\\\\
{\bf DOI:} 10.1134/S1063772917070022

\end{abstract}

\section{Introduction}

The EX Hydrae (EX Hya) binary system is one of the closest (at a distance of about 65 pc \cite{Beuermann2003}) and brightest ($9.6^m$ -- $14^m$) cataclysmic variables. It is an intermediate polar \cite{Warner}, that consists of a low mass M star (donor) and white dwarf (accretor). Intermediate polars are interacting binary systems in that material from the donor flows through the inner Lagrange point $L_1$ onto a white dwarf. In the process of the mass transfer, an accretion disk forms around the white dwarf. The white dwarfs in intermediate polars have magnetic fields strong enough to influence the flow in the inner parts of the accretion disk ($10^4$ -- $10^6$ G at the white dwarf surface). The interaction of the magnetic field and the plasma of the accretion disk results in the formation of a magnetosphere near the white dwarf. Accretion onto the white dwarf proceeds via accretion columns, that have a curtain-like shape and are mainly oriented along magnetic field lines \cite{Isakova2015}. The basic parameters of EX Hya are listed in the table 1 (see, e. g., \cite{Semena2014a}).

\begin{table}[th]
\renewcommand{\tabcolsep}{0.8cm}
\begin{center}
\caption{Parameters of EX Hya}
\begin{tabular}{lcc}
Parameter & Value & Reference \\
\hline
Accretor mass, $M_\text{a}$ & $0.79~M_{\odot}$ & \cite{Beuermann2008} \\
Accretor radius, $R_\text{a}$ & $0.7 \times 10^9~\text{cm}$ & \cite{Beuermann2008} \\
Inner radius of the disk, $R_\text{d}$ & $1.9 \times 10^9~\text{cm}$ & \cite{Siegel1989, Revnivtsev2012} \\
Magnetic field of accretor, $B_\text{a}$ & $8~\text{kG}$ & * \\
Accretion rate, $\dot{M}$ & $3 \times 10^{15}~\text{g/s}$ & ** \\
\hline
\end{tabular}
\end{center}
\begin{flushleft}
*~--- Assuming that the magnetosphere radius is equal to the inner
radius of the disk. \\
**~---  Assuming that the luminosity $L \approx G M_\text{a} \dot{M} \left(  1/R_\text{a} - 1 / R_\text{d} \right)$.
\end{flushleft}
\label{tab1}
\end{table}

Attempts to determine the geometric sizes and properties of the accretion columns in polars
and intermediate polars have been made in several studies. The basic methods used to determine these
parameters come down to spectral analyses of the X-ray radiation of these systems or analyses of their
periodic and stochastic variability. 

The simplest method for determining the area of a column is analysis of the soft X-ray radiation from the surface of the white dwarf, that is heated by hard X-rays from the accretion column \cite{haberl95}. Such an estimate for the intermediate polar EX Hya provided the limit on the column spot area $f=7.3_{-4.0}^{+29.3}\times10^{-4}$, where $f$ is the ratio of the area of the column and the entire surface area of the white dwarf \cite{evans07}. However, this method overestimates the area of the column base: depending on the geometry and height of the column, the area of the heated region at surface of the white dwarf can be a factor of a few larger than the area of contact of the column with the surface of the white dwarf. The shape of the hot spot and its location on the white dwarf surface are shown schematically in Fig. 1. The region of the white dwarf surface that is heated by the accretion column is shown in yellow.

\begin{figure}[t]  
\centering 
\includegraphics[width=0.55\textwidth, height=0.6\textwidth]{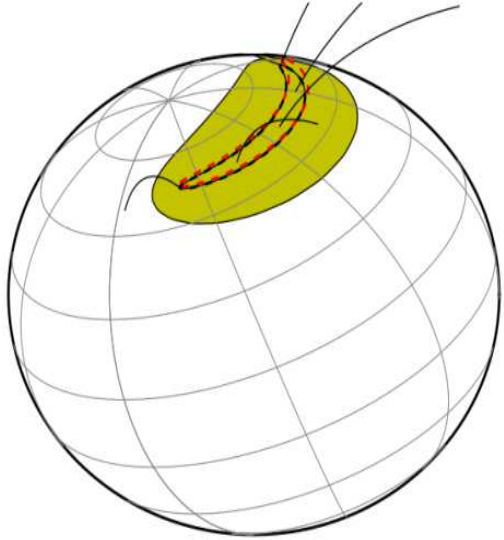}
\caption{ The shape and location of the hot spot on the surface of the white dwarf. The black contour marks the boundary of the contact area of the accretion stream and the surface of the white dwarf. The red dashed curve shows the position of the shock in the accretion flow. The portion of the white dwarf surface heated to high temperatures by hard X-ray emission from the hot zone of the accretion column (the zone bounded by the black contour and red dashed curve) is shaded yellow.}
\label{fig-bb}
\end{figure}

Another spectral method for determining the column area is analyzing the hard X-rays emitted by
the hot, optically thin plasma settling onto the white dwarf surface in the accretion column (see, e.g., \cite{aizu73,Lamb1979, langer81, canalle05,hayashi14a}). The area of the accretion column in EX Hya found in this way, $f<0.1_{-0.1}^{+3.4}\times10^{-2}$ \cite{hayashi14} , exceeds the area found using the other method above. However, the shape of the X-ray spectrum of the accretion column at $0.5$--$10$ keV (the energy range of the main modern X-ray missions) obtained via hydrodynamical
calculations of the flow in the column depends only weakly on the specific accretion rate. This results in appreciable uncertainty in estimates of the specific accretion rate and the area of the spot where the column is in contact with the white dwarf surface derived using this method (this can be seen from the fact that, in the case of the intermediate polar EX Hya, the latter method actually did not constrain the minimum area of the column).

Analysis of the energy fluxs in the emission lines of ionized elements \cite{blumenthal72} can also be used to estimate the material density in the column at the white dwarf surface, where a significant fraction of the energy of the falling material is released. However, this method requires high energy resolution and good data quality. An estimate of the column area in EX Hya obtained using this method is $f<0.02$ \cite{mauche01}. Analysis of a large number of emission lines in the X-ray for a fractional accretion column area $1.6\times10^{-4}$ was carried out in \cite{luna15}. This showed that, to explain the observed flux ratios in some lines, the chemical abundances in the accreted material must be significantly different from the solar abundances, and this method cannot provide
an unambiguous estimate of the material density at the base of the column, and hence of the column area.

The area of the accretion column can be estimated using temporal information on the object. EX Hya
is an eclipsing system with an inclination of about 80$^\circ$ \cite{hellier87}. Hard X-ray observations carried out by the RXTE Proportional Counter Array were used to fix the times of total X-ray eclipses in EX Hya \cite{Mukai1998}. Figure 2 shows schematically the appearance of the hot
spot on the white dwarf surface during the eclipses. The contours indicate the boundaries of the radiating region on the surface, based on the time of eclipse ingress and egress.

\begin{figure}[t]
\centering 
\includegraphics[width=0.54\textwidth, height=0.5\textwidth]{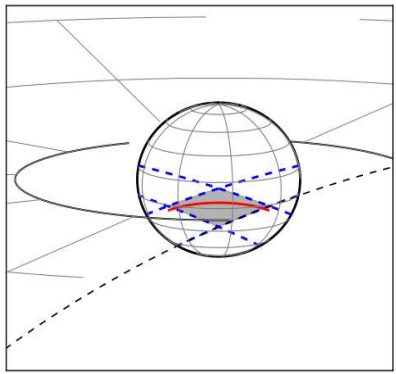}
\caption{Scheme of the geometry of the hot spot on the white dwarf surface during eclipses. The dotted curves mark the maximum size of the radiating zone (shaded in gray), based on the times of eclipse ingress and egress. The red curve marks the contact spot of the accretion column with the white dwarf surface, that determines the observed X-ray eclipse profile.} 
\label{fig-eal}
\end{figure}

The duration of the eclipse is determined by the time it takes for the companion star to cross the line of sight in the front of the white dwarf, while the durations of the eclipse ingress and egress (the time for the X-ray flux to decrease to zero and then to increase to its normal value) are determined by the size of the radiating zone. In the case of intermediate polars, it is known that the main source of X-ray radiation is hot material in the accretion column, close to the white dwarf surface. It was shown in \cite{Mukai1998} that the duration of the eclipse ingress is about 20 s, corresponding to the dimensions of the white dwarf. However, note that the duration of the eclipse ingress actually provides only a limit on the spatial size of the radiating spot, since a long, thin spot with a small area is allowed.

The area of the accretion column can be estimated based on the properties of the aperiodic variability
of the luminosity of the accretion column. This method is based on searching for a characteristic frequency in the power spectrum of the variability of the luminosity of the system, above that the variability amplitude begins to fall rapidly \cite{Semena2012}. The drop in the variability above this frequency is associated with the cooling time scale of the material of the accretion column. The rate of material flow in the accretion column is variable and contains the power spectrum of variability generated in the accretion disk (see, e.g., \cite{lyubarskii97}). The X-ray emission of an intermediate polar is due to cooling of optically thin plasma that is heated to temperatures  $\sim~10^{7}$~K during the passage of a shock by the white dwarf surface. At any given time, the X-ray luminosity of an intermediate polar is determined by the total amount of material that has passed through the shock over the typical cooling time. As a result, variability of the accretion rate should be suppressed at frequencies above the inverse cooling time of the material in the accretion column, that depends on the area of the accretion column. This method depends weakly on the specific cooling function of the material, the elemental abundances, and the orientation and shape of the column spot, hence providing an unbiased estimate of column area. The fractional area of the accretion column estimated in this way in \cite{Semena2014a} is $f<1.6\times10^{-4}$.

The results of most estimation methods applied to EX Hya lead to the conclusion that the area of
the accretion column does not exceed a fraction of a percent of the area of the white dwarf surface ($f<10^{-3}$).

In our current study, we have used two dimensional numerical simulation on a fairly detailed computational grid to investigate the structure of the accretion curtain in the inner part of the accretion disk. The computational results can be used to determine the thickness of the accretion curtain at the edge of the accretion disk, and thus, after taking into account the geometry of the magnetic field, to estimate the corresponding thickness of the curtain base at the accretor surface. In addition, comparison with observations provides information about the properties of the accretion disk.

The paper is organized as follows. In Section 2, we discuss the formulation of the problem and present
analytical estimates of the thickness of the accretion curtain. Section 3 describes our numerical model. The results of the numerical simulations are presented in Section 4, and discussed in Section 5. Our main conclusions are briefly summarized in the Conclusion.

\section{Formulation of the Problem}

We have assumed that the thickness of the accretion curtain at the surface of the accretor $\delta$ is related to its thickness at the base of the accretion disk $H_\text{m}$. A structure of the flow structure near the accretor is shown in Fig. 3. This shows the white dwarf (gray circle marked by 1), accretion disk (2), accretion curtains (3), and magnetic field lines (lines with arrows). $R_\text{a}$ is the radius of the accretor, $\delta$ the thickness of the base of the accretion curtain, $R_\text{m}$ the radius of the magnetosphere, $H_\text{m}$ the thickness of the accretion curtain at the boundary of the magnetosphere, and $H_\text{d}$ the half-thickness of the accretion disk.

\begin{figure}[t]
\centering 
\includegraphics[width=0.75\textwidth]{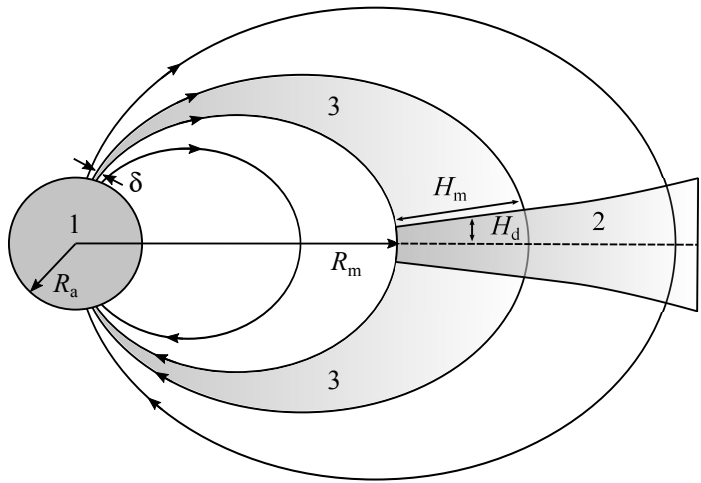}
\caption{Schematic view of the white dwarf magnetosphere, showing (1) the white dwarf with its magnetic field (lines with arrows), (2) the accretion disk, and (3) the accretion curtains. $\delta$ is the thickness of the accretion curtain at the white dwarf surface, $R_\text{a}$ the radius of white dwarf, $R_\text{m}$ the radius of the magnetosphere, $H_\text{m}$ the thickness of the accretion curtain at the boundary of the magnetosphere, and $H_\text{d}$ the half-thickness of the disk.
}
\label{fig-ms}
\end{figure}

Let make the reasonable assumption that the flow in the accretion curtains is determined completely by
the magnetic field of the white dwarf, so that the material inside the curtains flows along the magnetic field lines. We also assumed that the magnetic field of the white dwarf is dipolar \footnote{The magnetic fields of white dwarfs can also have a substantial quadrupolar component, and the character of the accretion can be more complex in this case (see, e.g., \cite{bycam1}). In particular, in addition to polar spots, equatorial accretion spots should form in this case. However, these are not observed in EX Hya. Therefore, we confined our consideration to the case of a purely dipolar magnetic field.}. The equation for dipolar magnetic field lines can be written
\begin{equation}\label{eq-mfl}
 R = R_0 \sin^2\theta,
\end{equation}
where $R_0$ is the distance from the stellar center where a given field line crosses the plane of the magnetic equator. We can write this equation for the two magnetic surfaces (a family of field lines \eqref{eq-mfl} with the same value of $R_0$ ) bounding the accretion curtain
\begin{equation}\label{eq-msurf}
 R_\text{a} = R_\text{m} \sin^2\theta_\text{a}, \quad
 R_\text{a} = \left( R_\text{m} + H_\text{m} \right) 
 \sin^2 ( \theta_\text{a} + \Delta\theta),
\end{equation}
where $\theta_\text{a}$ is the angular distance from the magnetic pole to the base of the curtain on the accretor surface, and $\Delta\theta < 0$ is the angle defining the thickness of the curtain base ($\delta = R_\text{a} |\Delta\theta|$). We find from the first relation in \eqref{eq-msurf} $\theta_\text{a} \approx 37.4^\circ$. Taking $\Delta\theta$ to be small, we obtain from the second relation in \eqref{eq-msurf}
\begin{equation}\label{eq-delta}
 H_\text{m} = 
 \frac{2\delta R_\text{m}}{R_\text{a} \tan\theta_\text{a} - 2\delta} \approx
 \frac{2\delta R_\text{m}}{R_\text{a} \tan\theta_\text{a}}.
\end{equation}
The area of the accretion spot at the white dwarf surface,
\begin{equation}\label{eq-Sa}
 S_\text{a} = 
 R_\text{a}^2 \sin\theta_\text{a} |\Delta\theta| \Delta \varphi = 
 \delta R_\text{a} \sin\theta_\text{a} \Delta \varphi
\end{equation}
is determined by the curtain opening angle $\Delta \varphi$.

To estimate the opening angles of the curtain, we performed 3D numerical simulations of the magnetosphere structure in the EX Hya system, applying the model described in \cite{Isakova2015}. In the computations, we adopted the magnetic field at the accretor surface $B_\text{a} = 8~\text{kG}$ and the inclination of the magnetic axis to the rotation axis of the white dwarf $30^\circ$. The results of computations are shown in Fig. 4, that presents isosurfaces of the logarithm of the density (color), magnetic field lines (lines with arrows), the rotation axis (solid blue line), and the magnetic axis (solid red line). Each accretion spot is part of an oval. The opening angles of the northern and southern accretion spots comprise approximately $\Delta \varphi = 170^\circ$.

\begin{figure}[t]
\centering 
\includegraphics[width=0.45\textwidth]{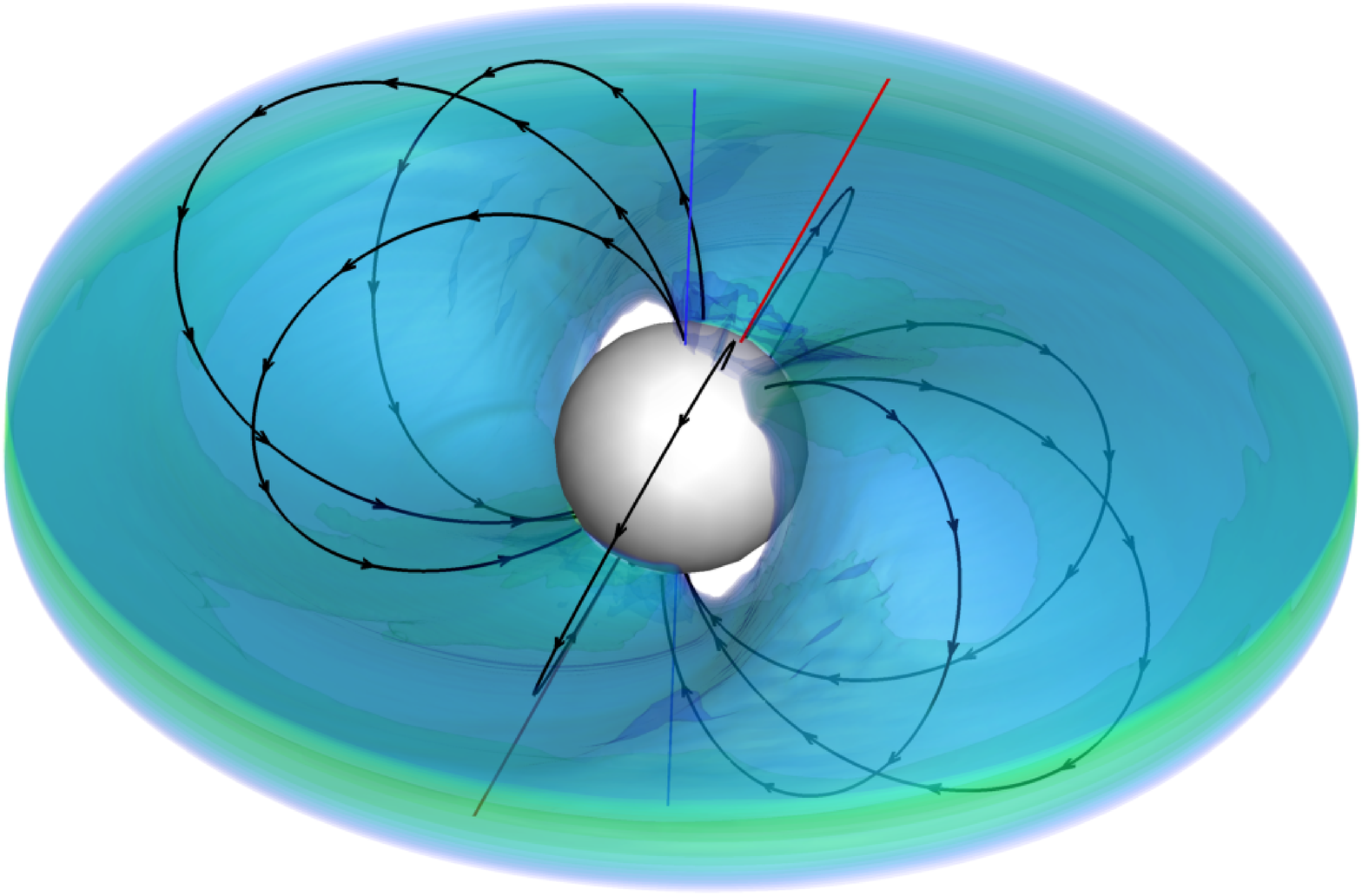}
\includegraphics[width=0.45\textwidth]{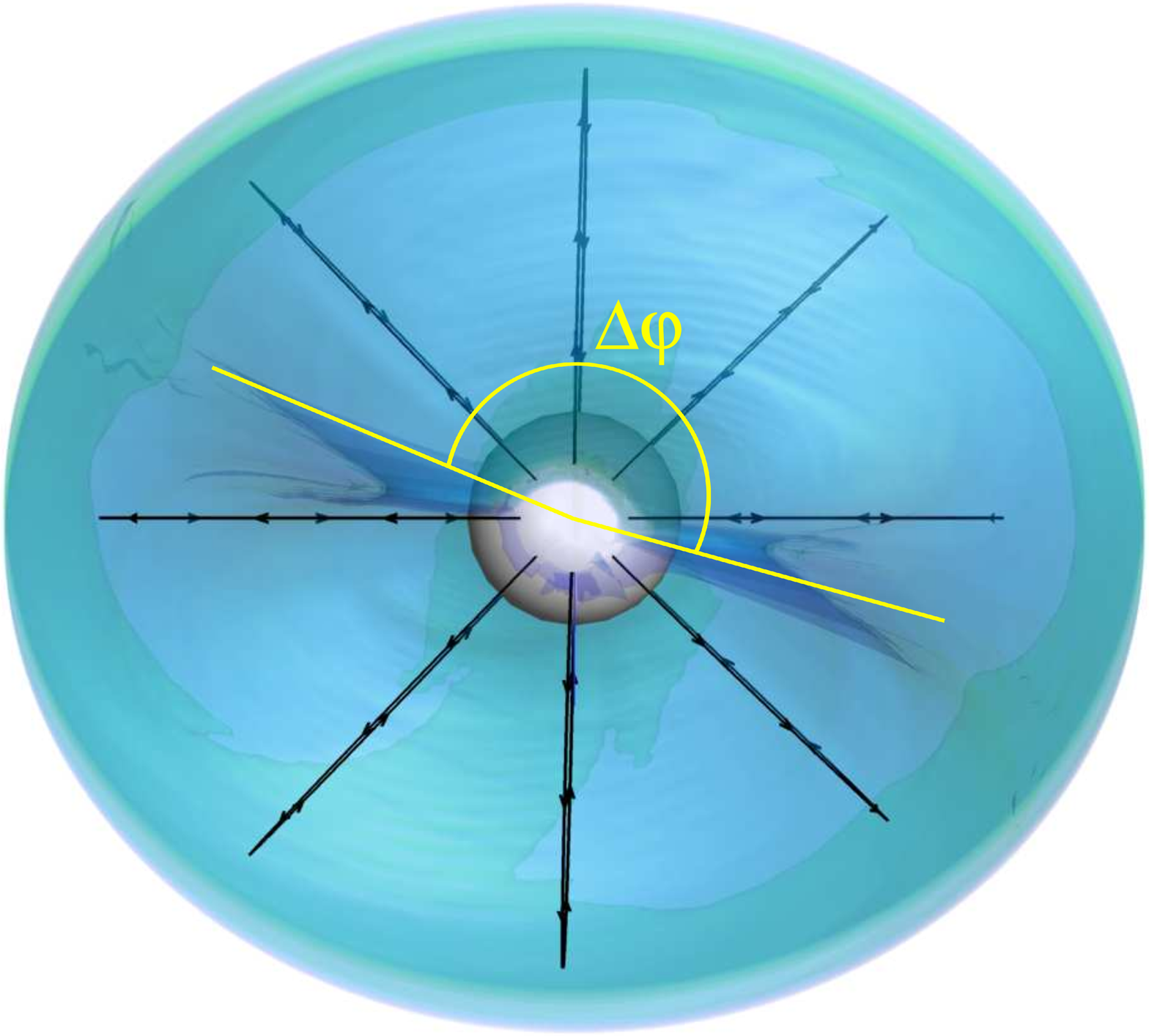}
\caption{3D flow structure in the magnetosphere. Shown are level surfaces of the logarithm of the density (color), magnetic field lines (lines with arrows), the rotation axis (blue line), and magnetic axis (red line). The diagram to the right shows the flow pattern as seen looking down on the northern magnetic pole of the star. The label $\Delta \varphi$ indicates the method used to determine the opening angle of the accretion curtain.}
\label{fig-3d}
\end{figure}

Taking into account the value for this angle obtained in the 3D computations, we obtain directly
from \eqref{eq-Sa} $\delta \approx 7.8~\text{km}$. It therefore follows from \eqref{eq-delta} that the thickness of the curtain at the edge of the disk is $H_\text{m} = 56~\text{km} = 0.003 R_\text{m}$.

We assume here that the thickness of the curtain at the boundary of the magnetosphere is mainly determined by the thickness of the accretion disk. From the condition of hydrostatic equilibrium in the vertical direction in the case of an isothermal disk, the disk half-thickness can be estimated by the corresponding scale height,
\begin{equation}\label{eq-Hd} 
 H_\text{d} = \sqrt{\frac{2 c_T^2 r^3}{G M_\text{a}}},
\end{equation}
where $G$ is the gravitational constant, $c_T$ the isothermal sound speed, and $r$ the distance from the center of the star to a given point in the equatorial plane of the disk. The thickness of the disk can differ from the value given by \eqref{eq-Hd} close to the boundary of the magnetosphere. However, for simplicity of our analysis, we used this expression to describe the disk half-thickness in the model. The specific value of $H_\text{d}$ will then be determined by the temperature $T$, that
was a free parameter in the computations.

\section{The Model}

We used the following system of equations to describe the flow structure in the region of formation of
the accretion curtain:
\begin{equation}\label{eq-rho1} 
 \pdiff{\rho}{t} + \nabla \cdot \left( \rho \vec{v} \right) = 0,
\end{equation}
\begin{equation}\label{eq-v1} 
 \pdiff{\vec{v}}{t} + \left( \vec{v} \cdot \nabla \right) \vec{v} =  
 -\frac{\nabla P}{\rho} - 
 \nabla \Phi - 
 \frac{\left(\vec{v} - \vec{u} \right)_{\perp}}{t_w},
\end{equation}
\begin{equation}\label{eq-s1}  
 \pdiff{s}{t} + 
 \left( \vec{v} \cdot \nabla \right) s = 0.
\end{equation}
Here, $\rho$ is the density, ${\bf v}$ the velocity, $P$ the pressure, $s$ the specific entropy, $\Phi$ the gravitational potential of the accretor, and $\vec{u}$ the velocity of the magnetic field
lines. The subscript $\perp$ denotes the component of a vector perpendicular to the magnetic field lines. We closed the system using the equation of state of an ideal gas:
\begin{equation}\label{eq-ig} 
 s = c_V \ln \left(P / \rho^{\gamma}\right),
\end{equation}
where $c_V$  is the specific heat capacity at constant volume, and the adiabatic index is $\gamma = 5/3$. Our model assumed that the accretor has a dipolar magnetic field with the vector magnetic field given by
\begin{equation}\label{eq-Bs}
 \vec{B}_{*} = 
 \fr{3(\vec{\mu} \cdot \vec{R}) \vec{R}}{R^5} -
 \fr{\vec{\mu}}{R^3},
\end{equation}
where $\vec{\mu}$ is the vector magnetic momentum of the star and $\vec{R}$ a vector drawn from the stellar center to a given point.

The above model is a modified version of the numerical model we used earlier to calculate the flow
structure in close binary systems with magnetic fields (see, e.g., \cite{zbSMF, zbbUFN2012, mcb-book, 
Isakova2015, Isakova2016}). The basic idea is that the dynamics of the plasma in a strong external magnetic field (corresponding to the situation in the white dwarf magnetosphere) is characterized by the relatively slow mean motion of particles along magnetic field lines, a drift due to gravity across the field lines, and Alfv{\`e}n and magnetoacoustic waves propagating at very high velocities. In this case, over a typical dynamical time scale, magnetohydrodynamical (MHD) waves have time to travel along the flow region (for example, an accretion column) multiple times. This makes it possible to investigate the averaged flow pattern, considering the effect of rapid pulsations by analogy with wave MHD turbulence.

The last term in the equation of motion \eqref{eq-v1} describes the force acting on the plasma due to the accretor magnetic field. This affects only the plasma velocity perpendicular to the magnetic field lines, $\vec{v}_{\perp}$. The justification for including this term is provided in the Appendix. The strong external magnetic field acts like an effective fluid, with that the plasma interacts. The last term in \eqref{eq-v1} can be interpreted as a friction force between the plasma and magnetic field
(more precisely, the magnetic field lines), whose form is similar to the friction force between the components in a plasma consisting of several types of particles (see, e.g., \cite{FrankKamenetsky1968}). The characteristic decay time for the transverse velocity is
\begin{equation}\label{eq-tw}
 t_w = \fr{4\pi\rho\eta_w}{B_{*}^2}.
\end{equation}
This value is determined by the wave dissipation of the magnetic field, that is characterized by the diffusion coefficient
\begin{equation}\label{eq-etaw}
 \eta_w = \alpha_w \fr{l_w B_{*}}{\sqrt{4\pi\rho}},
\end{equation}
where $l_w$ is the characteristic spatial scale of the wave pulsations, that can be estimated as the scale for inhomogeneities of the accretor magnetic field, $l_w = B_{*}/|\nabla B_{*}|$. The parameter $\alpha_w$, that is close to unity, describes the efficiency of the wave diffusion. We used the value $\alpha_w = 1/3$ in the model, corresponding to the isotropic wave MHD turbulence \cite{ZhilkinAIP2013}.

In our numerical model, we used the equation for the entropy \eqref{eq-s1} instead of an energy equation. It is known that this approximation can be used either in
the absence of shocks or when the amplitude of the shocks is small. In the case we have considered, the use of the entropy equation does not lead to the appearance of additional errors in the numerical solution, since the flow structure in the region of formation of the accretion curtain is essentially free of shocks.

Our aim was to study the structure of the accretion curtain near the inner edge of the disk. Therefore, to simplify the problem, we used an axisymmetric approximation with the direction of the disk rotation axis coincident with the axis of symmetry of the stellar magnetic field. We calculated the flow structure using a two dimensional numerical code that solved the system of equations \eqref{eq-rho1}--\eqref{eq-s1} in a spherical coordinate frame ($R$, $\theta$, $\varphi$) with its origin at the stellar center. A fairly dense grid ($256 \times 256$ cells) in a two dimensional computational domain was used for the solution: $R_{\min} \le R \le R_{\max}$, $\pi/2-\theta_0 \le \theta \le \pi/2 + \theta_0$ , where $R_{\min} = 0.5 R_\text{d}$, $R_{\max} = 3.5 R_\text{d}$, $R_\text{d}$ is the inner radius of the disk (see the table 1), and the angle $\theta_0$ was defined by the relation $2\sin\theta_0 = 1 - R_{\min}/R_{\max}$. The numerical code was based on a Godunov-type finite-difference scheme of high order of accuracy (see, e.g., \cite{ZhilkinMM2010}). The computations were carried out at the computer clusters of the Institute of Astronomy and the Joint Supercomputer Center of the Russian Academy of Sciences.

The accretion disk and corona were specified in the computational domain at the initial time. The density in the accretion disk was specified in accordance with the formula
\begin{equation}\label{eq-rho2}
 \rho = \rho_\text{d} e^{-z^2 / H_\text{d}^2},
\end{equation}
where $z = R \cos\theta$ is the distance from the equatorial plane to a given point in the disk. The value of $H_\text{d}$ was defined by \eqref{eq-Hd}. The density at the inner edge of the disk $\rho_\text{d}$ was determined based on the condition
\begin{equation}\label{eq-rho3}
 \frac{B_*^2}{8\pi} = \rho_\text{d} v_\text{K}^2,
\end{equation}
where $v_\text{K} = \sqrt{G M_\text{a} / R_\text{d}}$ is the corresponding Keplerian rotational velocity. We find from this last condition:
\begin{equation}\label{eq-rho4}
 \rho_\text{d} = \frac{\mu^2}{8\pi G M_\text{a} R_\text{d}^5}.
\end{equation}
The density of the material surrounding the disk was taken to be constant: $\rho_\text{c} = 10^{-6} \rho_\text{d}$. The temperature of the disk $T$, that determines the disk thickness $H_\text{d}$ in our model, was treated as a free parameter.

An important point on that our model is based is that the magnetic Reynolds number be small compared to unity, $\Rm \ll 1$ (see the Appendix). It is easy to verify that this condition is indeed satisfied in our initial formulation of the problem. The characteristic value of the magnetic Reynolds number is
\begin{equation}\label{eq-Rm1}
 \Rm = \frac{vL}{\eta},
\end{equation}
where $v$ is the characteristic velocity, $L$ the characteristic spatial scale, and $\eta$ the magnetic viscosity. Taking the Keplerian rotational speed at the inner
boundary of the disk as the typical velocity $v$ and the turbulent magnetic viscosity $\eta_w$, defined by \eqref{eq-etaw}, as the magnetic viscosity $\eta$, we find
\begin{equation}\label{eq-Rm2}
 \Rm = \frac{3}{\sqrt{2} \alpha_w} 
 \sqrt{\frac{\rho}{\rho_\text{d}}} 
 \frac{L}{R_\text{d}},
\end{equation}
where the estimated scale for inhomogeneity of the dipolar magnetic field is $l_w = R_\text{d} / 3$. For estimates in the accretion disk, we can take $\rho = 
\rho_\text{d}$ and $L = H_\text{d}$. This gives
\begin{equation}\label{eq-Rm3}
 \Rm = \frac{3}{\sqrt{2} \alpha_w} \frac{H_\text{d}}{R_\text{d}} \ll 1,
\end{equation}
since $H_\text{d} \ll R_\text{d}$. In the circumdisk corona, $\rho = \rho_\text{c} \ll \rho_\text{d}$ and $L = R_\text{d}$, so that
\begin{equation}\label{eq-Rm5}
 \Rm = \frac{3}{\sqrt{2} \alpha_w} 
 \sqrt{\frac{\rho_\text{c}}{\rho_\text{d}}} \ll 1.
\end{equation}
Thus, the small Reynolds number approximation, $\Rm \ll 1$, that is the basis of our model, is fairly well justified.

\section{Results}

To study the influence of the disk thickness on the thickness of the accretion curtain that forms at the boundary of the magnetosphere, we performed numerical computations of the flow structure for three disk temperatures $T$: $10^3$~K (Model 1), $10^4$~K (Model 2), and $10^5$~K (Model 3). In all cases, the computations were continued until a steady-state flow regime was established. The typical time required for this was about 15--20 Keplerian rotation periods at the inner radius of the disk $R_\text{d}$.

\begin{figure}[t]  
\centering 
\includegraphics[width=0.75\textwidth]{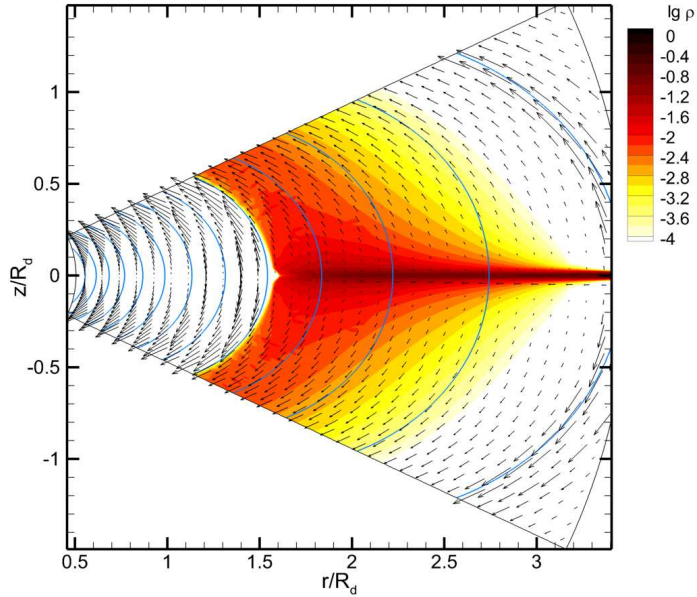}
\caption{Distribution of the density (color scale) and velocity (arrows) in the meridional ($\varphi = \text{const}$) plane of Model 1 (disk temperature $T = 10^3$~K). The cylindrical coordinates are defined by $r = R\sin\theta$, $z = R\cos\theta$. The density is shown in units of $\rho_\text{d}$. The curved lines correspond to the stellar magnetic field lines.} 
\label{fig-c3}
\end{figure}

\begin{figure}[t]  
\centering 
\includegraphics[width=0.75\textwidth]{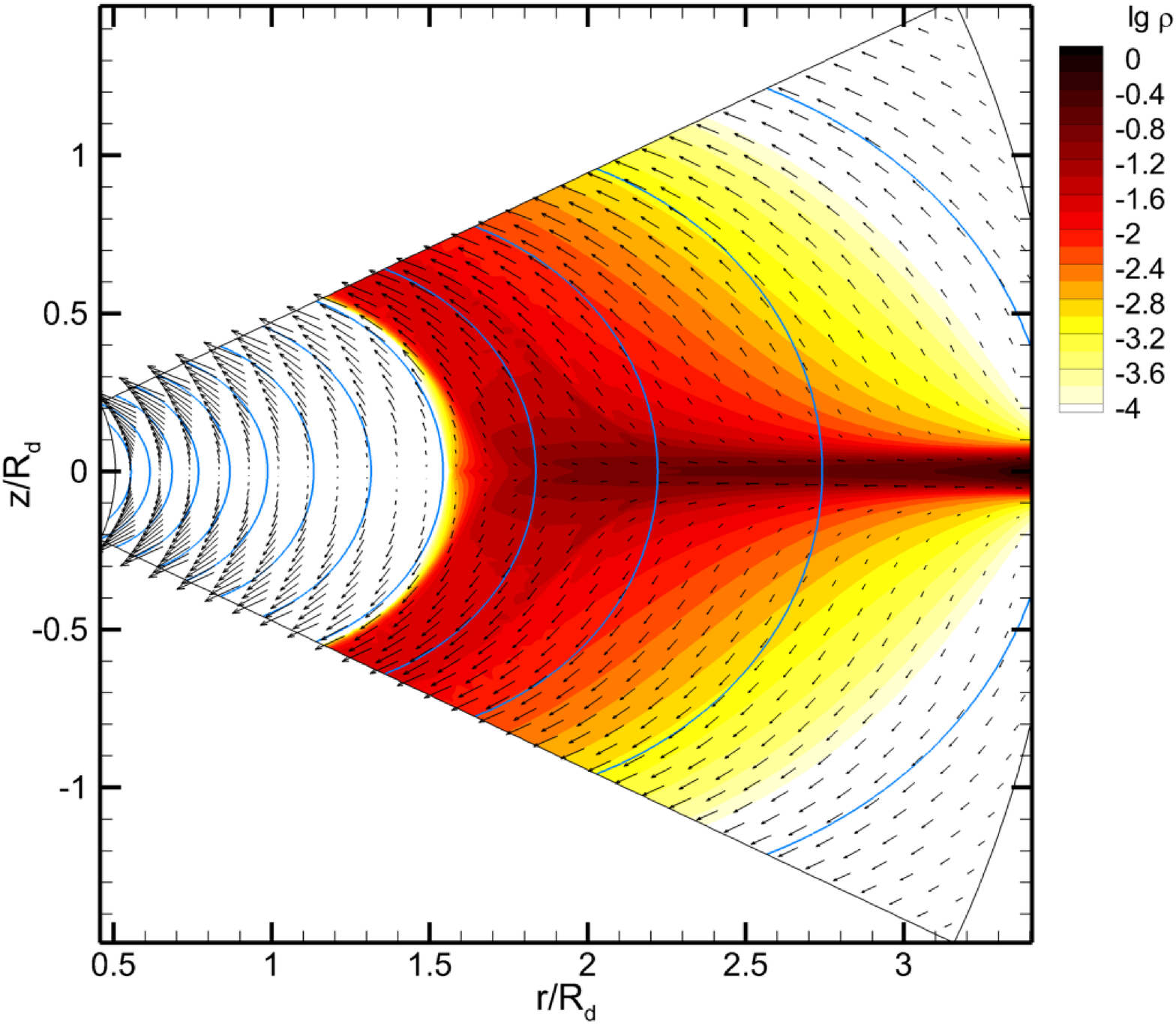}
\caption{Same as Fig. 5 for Model 2 (disk temperature $T = 10^4$~K).}
\label{fig-c4}
\end{figure}

\begin{figure}[t]  
\centering 
\includegraphics[width=0.75\textwidth]{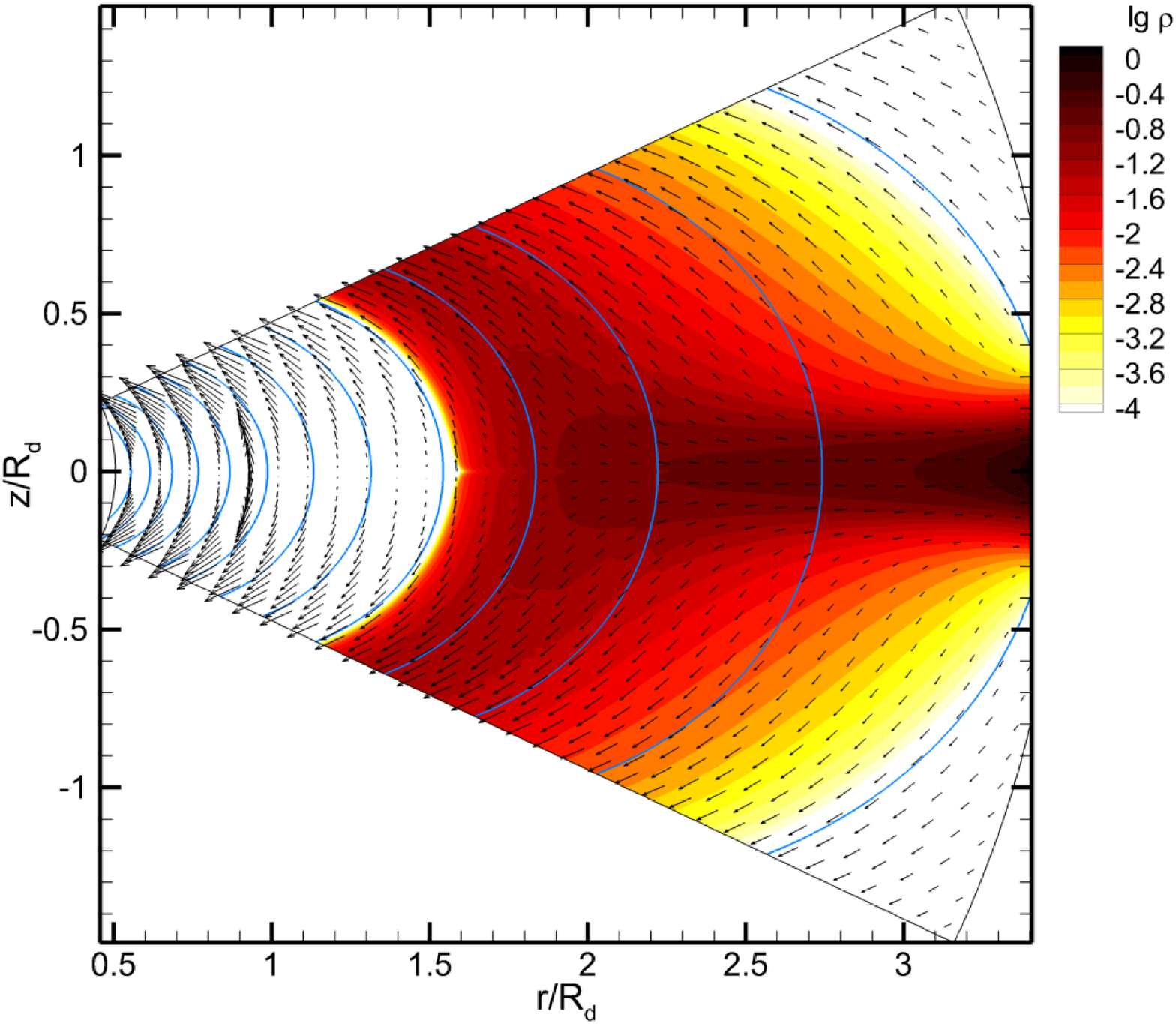}
\caption{Same as Fig. 5 for Model 3 (disk temperature $T = 10^5$~K).}
\label{fig-c5}
\end{figure}

The flow structure found in the computations is shown in Figs. 5, 6, and 7, that show the density distribution (color-scale) and velocity (arrows) in a meridional plane ($\varphi = \text{const}$) in cylindrical coordinates, $r = R\sin\theta$, $z = R\cos\theta$. The magnetic field lines are shown by solid curves. The spatial coordinates are normalized to the inner radius of the disk $R_\text{d}$, and the density is expressed in the units of $\rho_\text{d}$ \eqref{eq-rho4}.

A magnetosphere forms close to the accretor surface, and the accretion has a pronounced column-like character. The characteristic radius of the magnetosphere is about $1.5 R_\text{d}$ in all the models, corresponding to approximately four white dwarf radii. The radius of the magnetosphere turned out to be slightly larger than the observed inner radius of the disk. This is due to the fact that we set a predefined density at the inner radius of the disk in the initial conditions. As a result of the formation of the accretion flow, the density at the inner radius of the disk drops, and the boundary of the magnetosphere shifts to the right of the star. However, we emphasize that this is not important for our study, since the characteristic thickness of the accretion curtain changes only slightly.

A vacuum region forms inside the magnetosphere, that is practically devoid of material. The plasma in this region moves along magnetic field lines. Motion in the transverse direction is hindered by the electromagnetic force, that is described by the last term in Eq.~\eqref{eq-v1} in our model.

\begin{figure}[t]  
\centering 
\includegraphics[width=0.75\textwidth]{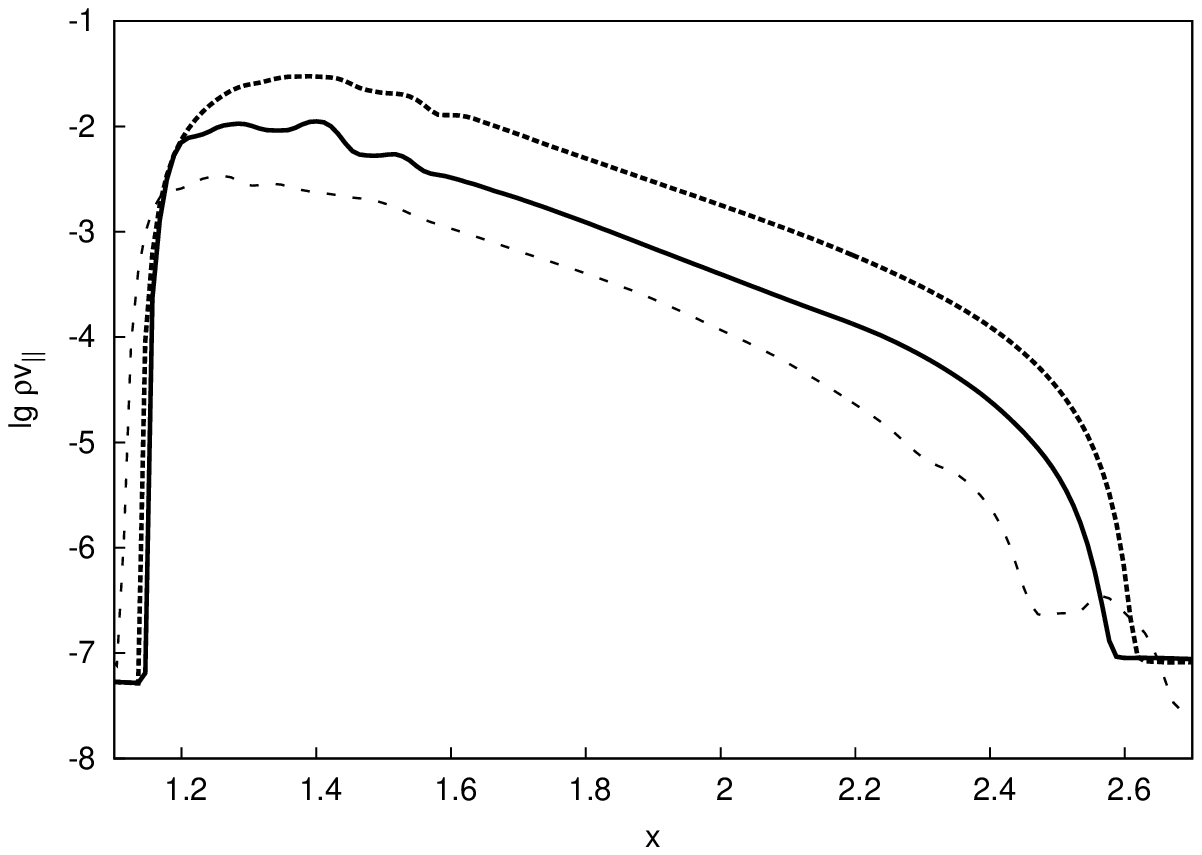}
\caption{Distribution of the logarithm of the mass flux density over the cross section of the accretion flow for three values of the temperature of the accretion disk: $10^3$~K (long-dashed curve), $10^4$~K (solid curve), and $10^5$~K (short-dashed curve).}
\label{fig-rhov}
\end{figure}

The action of the electromagnetic force on the disk plasma initially leads to the braking of the rotation in the inner regions of the disk. Farther, the material that has lost angular momentum begins to fall onto the star. Since the electromagnetic force hinders the motion of the plasma across the magnetic field lines, the infall (accretion) of the plasma occurs in the longitudinal direction relative to the magnetic field. This leads to the formation of accretion curtains, with corresponding regions of energy release (accretion spots) in the vicinity of the magnetic poles of the star. The figures show that, with increasing disk thickness (temperature), the thickness and characteristic density of the accretion curtains also increase. More accurate estimates of the thickness of the curtains can be obtained from an analysis of the distribution of the mass flux density over the cross section of the accretion flow, shown in Fig. 8. The coordinate $x$ corresponds to the radial coordinate $R/R_\text{d}$ along the direction defined by the angle $\theta = \pi/2 - \theta_0$. The mass flux density is expressed in units of $\rho_\text{d} \sqrt{G M_\text{a} / R_\text{d}}$. The component of the vector $\rho\vec{v}$ along the magnetic field lines is shown. Figure 8 shows that the mass flux density in the accretion curtain can be approximated well by the exponential function
\begin{equation}\label{eq-rhov}
 \rho v_\parallel = A e^{-x/H},
\end{equation}
where $H$ is the scale of inhomogeneity of the flow, that can be identified with the characteristic thickness of the curtain. Fitting formula \eqref{eq-rhov} to the numerical values of the flux density shown in Fig. 8 yields $H = 0.155 R_\text{d}$ for Model 1, $H = 0.193 R_\text{d}$d for Model 2, and $H = 0.208 R_\text{d}$ for Model 3. These values show that there is some dependence $H_\text{m}(H_\text{d})$, but it is fairly weak. In all the models, the thickness of the accretion curtain at the magnetosphere boundary is about 20\% of the inner disk radius.

\section{Discussion} 

The results of the computations described in the previous section are in good agreement with the results obtained in other studies (see, e.g., \cite{Romanova2002}). However, we consider them to be unsatisfactory from the point of view of observations. The main conclusion following from an analysis of the results is that, in
the stated formulation of the problem, the accretion curtain that is formed is too thick. Thus, in order to explain the observational data (a thin accretion curtain), we must invoke some other ideas.

One of the main assumptions underlying our model is that the magnetic field of the star penetrates fully into the accretion disk. In this case, the characteristic thickness of the accretion curtain $H_\text{m}$ should be defined by the scale for inhomogeneity of the magnetic field $l_w$. Close to the boundary of the magnetosphere, this scale can be estimated as $l_w = R_\text{d} / 3$. This simple estimate is in fairly good agreement with the thickness of the accretion curtain (fraction of the inner disk radius) obtained from the numerical simulations. However, this assumption may be not satisfied, since the accretion disk is formed by material of the donor envelope. In the case of a conducting plasma, the disk will be ideal diamagnetic. Such an accretion disk will be squeezed by the magnetic field of the accretor. The magnetic field will be virtually absent in the disk iteself.

\begin{figure}[t]  
\centering 
\includegraphics[width=0.75\textwidth]{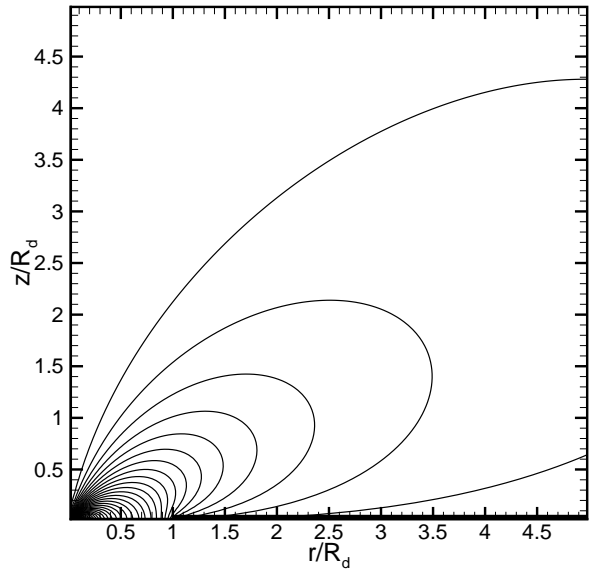}
\caption{Structure of the magnetic field of a star with an infinitesimally thin, diamagnetic accretion disk in the $r-z$ plane. The accretion disk is shown by the thick line along the $r$ axis.}
\label{fig-mf}
\end{figure}

The structure of the dipolar magnetic field of the star in the presence of an ideally conducting thin disk was first considered analytically in \cite{Aly1980, Kundt1980}. In the axisymmetric case, the structure of the magnetic field is given by the vector potential ($\vec{B} = \rotor\vec{A}$), that has the form
\begin{equation}\label{eq-A1}
 A_\varphi = \frac{2\mu\sin\theta}{\pi R_\text{d}^2} \left(
 X - \frac{\cos^2\theta}{\xi^2 X} + 
 \frac{\sin^2\theta}{\xi^2} \arctan X
 \right),
\end{equation}
where
\begin{equation}\label{eq-A2}
 \xi = \frac{R}{R_\text{d}}, \quad
 X = \frac{\sqrt{2} |\cos\theta|}{\sqrt{\xi^2 - 1 + Y}}, \quad
 Y = \sqrt{(\xi^2 - 1)^2 + 4 \xi^2 \cos^2\theta}.
\end{equation}

The structure of the magnetic field (in the plane of the cylindrical coordinates $r = R \sin\theta$ and $z = R \cos\theta$) described by the vector potential \eqref{eq-A1} is shown in Fig. 9; the infinitely thin accretion disk is shown by the thick line along the $r$ axis. The presence of a conducting disk leads to a strong distortion of the original stellar magnetic field. Close to the star ($R \to 0$), the field remains dipolar, while the magnetic field acquires a quadrupole character at large distances ($R \to \infty$), $B \propto R^{-4}$.

The formation of the accretion curtain has a completely different character in such a model, where the accretion curtain can form only in a relatively thin layer where the stellar magnetic field interacts with the disk plasma. The penetration of the plasma of a diamagnetic disk into the stellar magnetosphere can occur due to either diffusion of the magnetic field or the development of instabilities at the boundary of the magnetosphere. In any case, the characteristic thickness of the accretion curtain will be determined by the characteristic scale of the penetration of the field into the plasma.

Let us estimate the thickness of the accretion curtain in the case when the magnetic field penetrates into the plasma due to diffusion. The thickness of the curtain will then be equal to the thickness of the diffusion layer, $H_\text{m} \approx \sqrt{\eta \tau}$, where $\eta$ is the corresponding coefficient of magnetic viscosity and $\tau$ a characteristic time scale. Taking $\tau = \omega_\text{K}^{-1}$ and taking $\eta$ to be the Bohm diffusion coefficient (see, e.g., \cite{Chen}),
\begin{equation}\label{eq-etaB}
 \eta_\text{B} = \frac{1}{16} \frac{ckT}{eB},
\end{equation}
we obtain
\begin{equation}\label{eq-hm2}
 H_\text{m} \approx
 \frac{c_s}{\omega_\text{K}} 
 \sqrt{\frac{\omega_\text{K}}{\omega_\text{c}}} \approx
 2.5 \times 10^{-4} H_\text{d}.
\end{equation}
In these formulas, $c$ is the speed of light, $k$ Boltzmann constant, $e$ the elementary charge, $c_s$ the sound speed, and $\omega_\text{c}$ the cyclotron frequency for protons. Recall that the physical nature of Bohm diffusion is associated with the development of drift-dissipative instabilities, that lead to further turbulization of the plasma.

As an example, we will take the disk temperature to be $T = 10^4~\text{K}$. The relations \eqref{eq-hm2} provide the estimation $H_\text{m} \approx 2400$~cm or $H_\text{m} \approx 1.2 \times 10^{-6} R_\text{d}$. The diffusion layer (and, therefore, the accretion curtain) is very thin in this case. This value satisfies the observational data with a large reserve (two to three orders of magnitude). In future, we intend to modify our numerical model to take into account the effect of a diamagnetic disk.

\section{Conclusion}

We have proposed a two dimensional numerical model describing the flow structure in the magnetosphere of a white dwarf in an intermediate polar in an axisymmetric approximation. It was assumed that the white dwarf has a dipolar magnetic field whose symmetry axis coincides with the disk axis of symmetry. The model is based on the approximation of a poorly conducting plasma in a strong external magnetic field. In this case, the effect of the magnetic field reduces to taking into account the electromagnetic force in the equation of motion, that has a form equivalent to a frictional force. The approximation of small magnetic Reynolds number is satisfied with
a good reserve in the problem considered, if it is assumed that the plasma conductivity is determined by wave MHD turbulence.

We have applied this model to the EX Hya system, focusing mainly on the dependence of the thickness of the accretion curtains on the thickness of the accretion disk. The results of our numerical simulations show the formation of a flow structure with well known features close to the white dwarf. The accretion has a pronounced column-like character, and the accretion columns formed are not closed, and have the shape of curtains. A vacuum region that is virtually devoid of material forms inside the magnetosphere. The plasma in this region moves along the magnetic field lines.

The simulation results show that the thickness of the accretion curtains depends only weakly on the thickness of the accretion disk. However, the calculated and observed thicknesses of the accretion curtains do not agree, with the observed curtain thickness being considerably smaller. The main reason for the formation of relatively thick accretion curtains in our model is the assumption that the stellar magnetic field penetrate fully into the disk plasma. A more attractive option may be a diamagnetic disk that fully or partially shields the stellar magnetic field. In this case, the thickness of the accretion curtain should be determined by the thickness of the diffusion layer at the boundary of the magnetosphere. Simple estimates show that the resulting values are in good agreement with the observational data. In our opinion, this assumption is quite natural, since the accretion disks in intermediate polars are formed by material from the donor envelope. In future, we plan to modify our model to take
into account the effect of a diamagnetic disk.

\section*{Acknowledgments}

We thank E.P. Kurbatov for useful discussions. P.B.I. was supported by the Russian Foundation for Basic Research (project 14-29-06059). A.G.Z. and D.V.B. were supported by the Russian Science Foundation (project 15-12-30038). A.N.S. and M.G.R. were supported by the Russian Science Foundation (project 14-12-01287).

\section*{Appendix}

The total magnetic field in the plasma $\vec{B}$ can be represented as a superposition of the stellar magnetic field $\vec{B}_*$ and the magnetic field induced by electric currents in the plasma $\vec{b}$,  $\vec{B} = \vec{B}_* + \vec{b}$. Since the background magnetic field $\vec{B}_*$ is stationary ($\spdiff{\vec{B}_*}{t} = 0$) and potential ($\rotor\vec{B}_* = 0$) in our case, the induction equation can be written
\begin{equation}\label{eq-b1}
 \pdiff{\vec{b}}{t} = \rotor\left(
 \vec{v} \times \vec{b} + 
 \vec{v} \times \vec{B}_* -
 \eta \rotor\vec{b} 
 \right),
\end{equation}
where $\eta$ is the coefficient of magnetic viscosity. In the limit of small magnetic Reynolds numbers, $\Rm \ll 1$, the magnetic field will rapidly decay ($\spdiff{\vec{b}}{t} \to 0$), so that its magnitude will be small compared to the magnitude of the background field, $b \ll B_*$ \cite{Braginsky1959}. Comparing the two last terms on the right-hand side of \eqref{eq-b1}, we can see that $b \approx \Rm B_*$. This means that the following relation will be valid in the stationary regime:
\begin{equation}\label{eq-b2}
 \vec{v} \times \vec{B}_* - \eta \rotor\vec{b} = 
 -c\, \vec{E} = c\, \nabla\phi,
\end{equation}
where $\phi$ is the scalar potential of the electric field $\vec{E}$. Let us introduce for convenience the velocity of the magnetic field lines $\vec{u}$, that is defined by the expression
\begin{equation}\label{eq-u1}
 \vec{E} = -\frac{1}{c} \left( \vec{u} \times \vec{B}_* \right).
\end{equation}
We then find from \eqref{eq-b2} and \eqref{eq-u1}
\begin{equation}\label{eq-j1}
 \rotor\vec{b} = \frac{1}{\eta} 
 \left[ 
 \left( \vec{v} - \vec{u} \right) \times \vec{B}_* 
 \right].
\end{equation}

In a magnetohydrodynamical approximation, the electromagnetic force in the equation of motion is \cite{Landau8}
\begin{equation}\label{eq-f1}
 \vec{f} = 
 -\frac{\vec{B} \times \rotor\vec{B}}{4\pi\rho} = 
 -\frac{\vec{b} \times \rotor\vec{b}}{4\pi\rho}  
 -\frac{\vec{B}_* \times \rotor\vec{b}}{4\pi\rho} \approx 
 -\frac{\vec{B}_* \times \rotor\vec{b}}{4\pi\rho},
\end{equation}
where we have assumed $b \ll B_*$. Inserting \eqref{eq-j1} into this last expression yields
\begin{equation}\label{eq-f2}
 \vec{f} = -\frac{B_*^2}{4\pi\rho\eta} 
 \left( \vec{v} - \vec{u} \right)_\perp.
\end{equation}
We used this expression for the electromagnetic force in our model in the equation of motion \eqref{eq-v1}. We took $\eta_w$, defined by wave MHD turbulence, to be the magnetic viscosity.

The velocity of the magnetic field lines $\vec{u}$ can be found as follows. Neglecting the coordinate dependence of the coefficient of magnetic viscosity and calculating the divergence of both sides of Eq. \eqref{eq-b2}, we get the Poisson equation for the scalar potential of the electric field:
\begin{equation}\label{eq-phi1}
 \nabla^2 \phi = \frac{1}{c} \vec{B}_* \cdot \rotor\vec{v}.
\end{equation}
When calculating the expression on the right-hand side of this equation, we used the condition that the magnetic field is potential, $\rotor\vec{B}_* = 0$. If this equation is solved and the potential $\phi$ is found, the transverse component of the magnetic field lines can be derived from \eqref{eq-u1}:
\begin{equation}\label{eq-u2}
 \vec{u}_\perp = c\, \frac{\vec{B}_* \times \nabla\phi}{B_*^2}.
\end{equation}

Let us estimate the influence of this effect on the dynamics of the accretion disk. In an axisymmetric approximation, in cylindrycal coordinates ($r$, $\varphi$, $z$),
\begin{equation}\label{eq-brotv1}
 \vec{B}_* \cdot \rotor\vec{v} = 
 \frac{B_{*z}}{r} \pdiff{}{r} \left( r v_\varphi \right) - 
 B_{*r} \pdiff{v_\varphi}{z}.
\end{equation}
In the accretion disk, we can approximately specify $v_\varphi = v_\text{K}$, $B_{*r} = 0$, $B_{*z} = -\mu/r^3$. Therefore,
\begin{equation}\label{eq-brotv2}
 \vec{B}_* \cdot \rotor\vec{v} = -\frac{\mu}{2} \sqrt{G M_\text{a}} r^{-9/2}.
\end{equation}
Since the source term \eqref{eq-brotv2} in the approximation used does not depend on the vertical coordinate $z$, the derivatives of the potential $\phi$ with respect to $z$ in the Laplace operator can be ignored in the Poisson Eq. \eqref{eq-phi1}. As a result, the Poisson equation can be written
\begin{equation}\label{eq-phi2}
 \frac{1}{r} \diff{}{r} \left( r \diff{\phi}{r} \right) = 
 -\frac{\mu}{2c} \sqrt{G M_\text{a}} r^{-9/2}.
\end{equation}
The solution of this equation is
\begin{equation}\label{eq-phi3}
 \phi = -\frac{2\mu}{25c} \sqrt{G M_\text{a}} r^{-5/2}.
\end{equation}

The velocity of the magnetic field lines \eqref{eq-u2} is
\begin{equation}\label{eq-u3}
 u_\perp = u_\varphi = 
 c \frac{B_{*z}}{B_*^2} \diff{\phi}{r} = 
 -\frac{v_\text{K}}{5}.
\end{equation}
The difference in the velocities that enters the expression for the electromagnetic force {eq-f2}) is
\begin{equation}\label{eq-u4}
 v_\varphi - u_\varphi = \frac{6}{5} v_\text{K}.
\end{equation}
Thus, the effect of the scalar potential in the accretion disk is to provide a correction to the electromagnetic force of about 20\%. Therefore, it must be taken into
account.

In our numerical model, the Poisson equation \eqref{eq-phi1} for the potential $\varphi$ is solved by means of an expansion in a series of Legendre polynomials:
\begin{equation}\label{eq-phi4}
 \phi(R, \theta) = \sum\limits_{n = 0}^{\infty} \phi_n(R) P_n(\cos\theta).
\end{equation}
The coefficients $\phi_n(R)$ satisfy the equation
\begin{equation}\label{eq-phi5}
 \frac{1}{R^2} \diff{}{R} \left( R^2 \diff{\phi_n}{R} \right) - 
 \frac{n(n+1)}{R^2} \phi_n = q_n,
\end{equation}
where $q_n(R)$ are the coefficients in the series expansion in Legendre polynomials of the right-hand side of the Poisson equation \eqref{eq-phi1}. In our approach, this equation was solved numerically using the tridiagonal matrix algorithm, with the following boundary conditions:
\begin{equation}\label{eq-phi6}
 \left. \diff{\phi_n}{R} \right|_{R=R_{\min}} = 0, \quad 
 \left. \phi_n \right|_{R=R_{\max}} = \phi_n^*,
\end{equation}
where $R_{\min}$ and $R_{\max}$ are the radii of the inner and outer boundaries of the computational domain in the coordinate $R$. To determine $\phi_n^*$, we write the explicit solution of \eqref{eq-phi5}:
\begin{equation}\label{eq-phi8}
 \phi_n(R) = -\frac{1}{2n+1} \left[
 \frac{1}{R^{n+1}}
 \int\limits_{0}^{R} \xi^{n+2} q_n(\xi) d\xi + 
 R^{n}
 \int\limits_{R}^{\infty} \frac{q_n(\xi)}{\xi^{n-1}} d\xi
 \right].
\end{equation}
Assuming that $q_n = 0$ for $R < R_{\min}$ , while q n is a specified function of the radius $R$ for $R > R_{\max}$, we can rewrite \eqref{eq-phi8} at the outer boundary $R = R_{\max}$,
\begin{equation}\label{eq-phi9}
 \phi_n^* = 
 -\frac{Q_n}{R_{\max}^{n+1}} 
 - A_n R_{\max}^{n},
\end{equation}
where
\begin{equation}\label{eq-Qn1}
 Q_n = \frac{1}{2n+1} \int\limits_{R_{\min}}^{R_{\max}}
 R^{n+2} q_n(R) dR,
\end{equation}
\begin{equation}\label{eq-An1}
 A_n = \frac{1}{2n+1}
 \int\limits_{R_{\max}}^{\infty} \frac{q_n(R)}{R^{n-1}} dR.
\end{equation}

The values of the multipole moments $Q_n$ vary with time, since the distribution of the sources $q_n(R)$ inside the interval $R_{\min} \le R \le R_{\max}$ changes. The values of the coefficients $A_n$ do not change, since the magnetic field and velocity in the outer region can be considered to be constant. The application of the condition $q_n = 0$ in the domain $R < R_{\min}$ is due to the fact that the plasma velocity in the magnetosphere must be along the magnetic field close to the surface
of the star. As a consequence, $\vec{v} \times \vec{B}_* = 0$ and $\vec{B}_* \cdot \rotor\vec{v} = 0$.

\begin{figure}[t]  
\centering 
\includegraphics[width=0.49\textwidth]{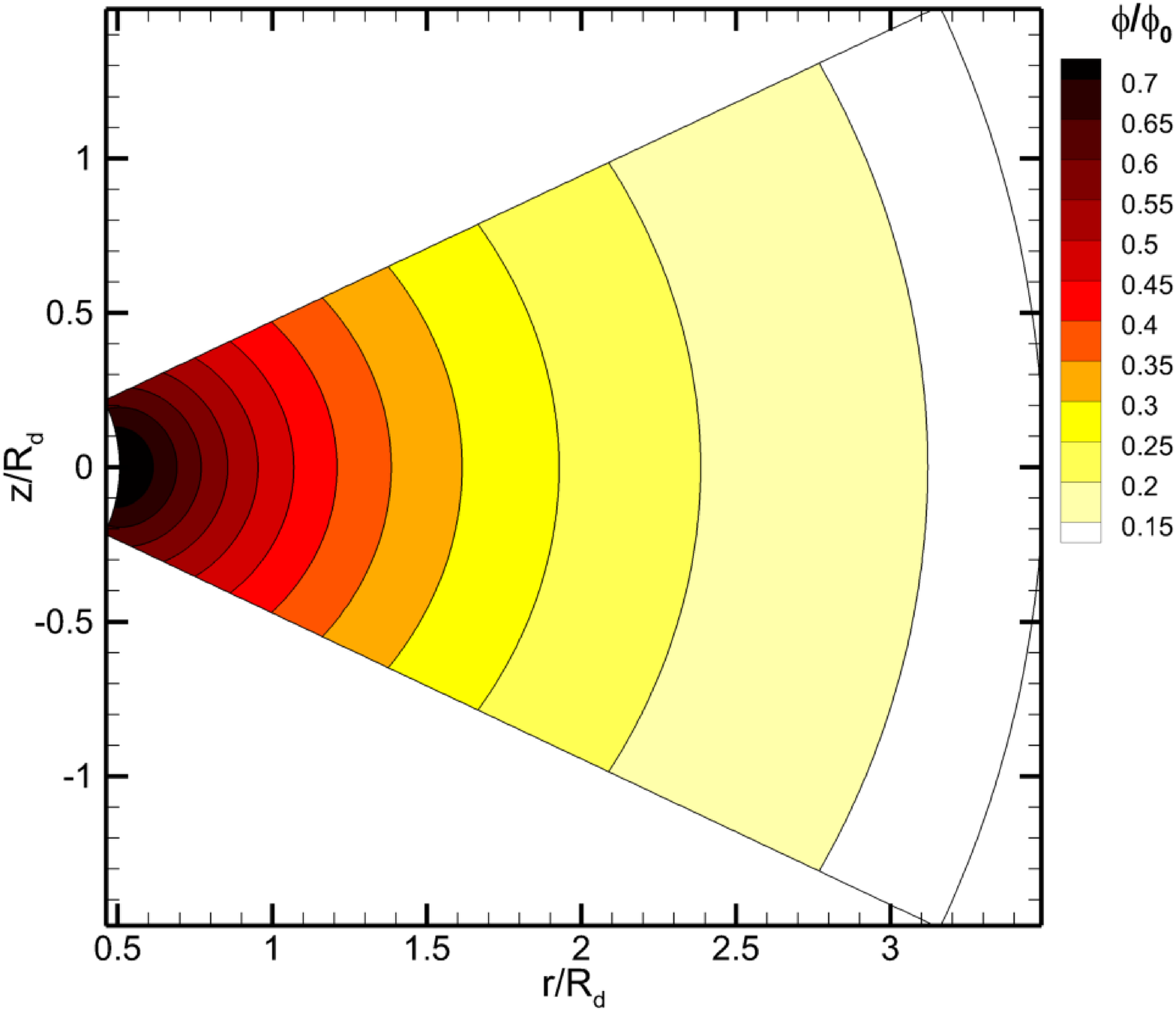}
\includegraphics[width=0.49\textwidth]{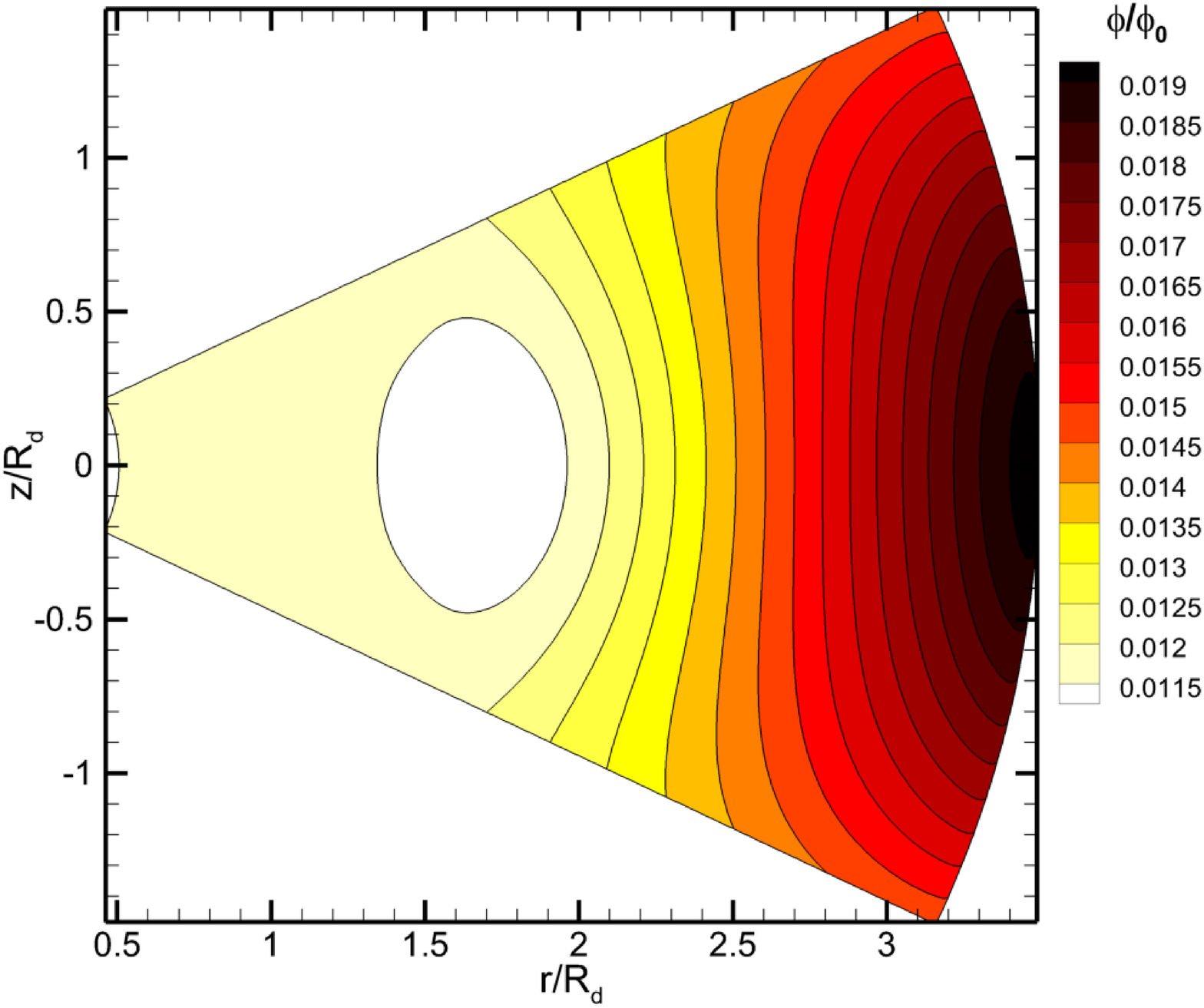}
\caption{Distribution of the scalar potential $\phi$ in Model 2 (temperature $T = 10^4$~K) at the initial time (left panel) and in the steady-state regime (right panel). The scaling factor is $\phi_0 = \sqrt{4\pi\rho_\text{d}} GM_\text{a}/c$.}
\label{fig-phi}
\end{figure}

\begin{figure}[t]  
\centering 
\includegraphics[width=0.49\textwidth]{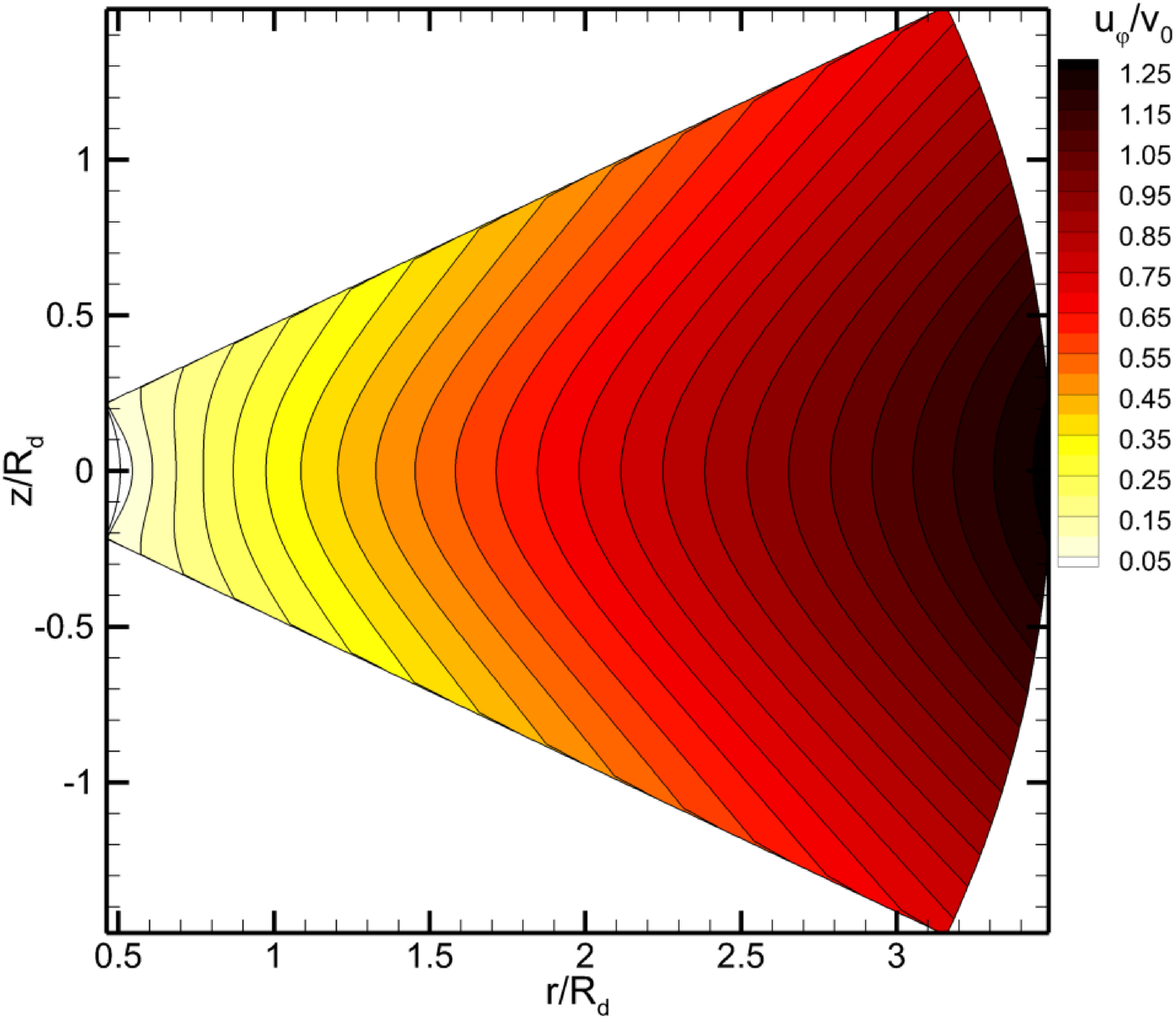}
\includegraphics[width=0.49\textwidth]{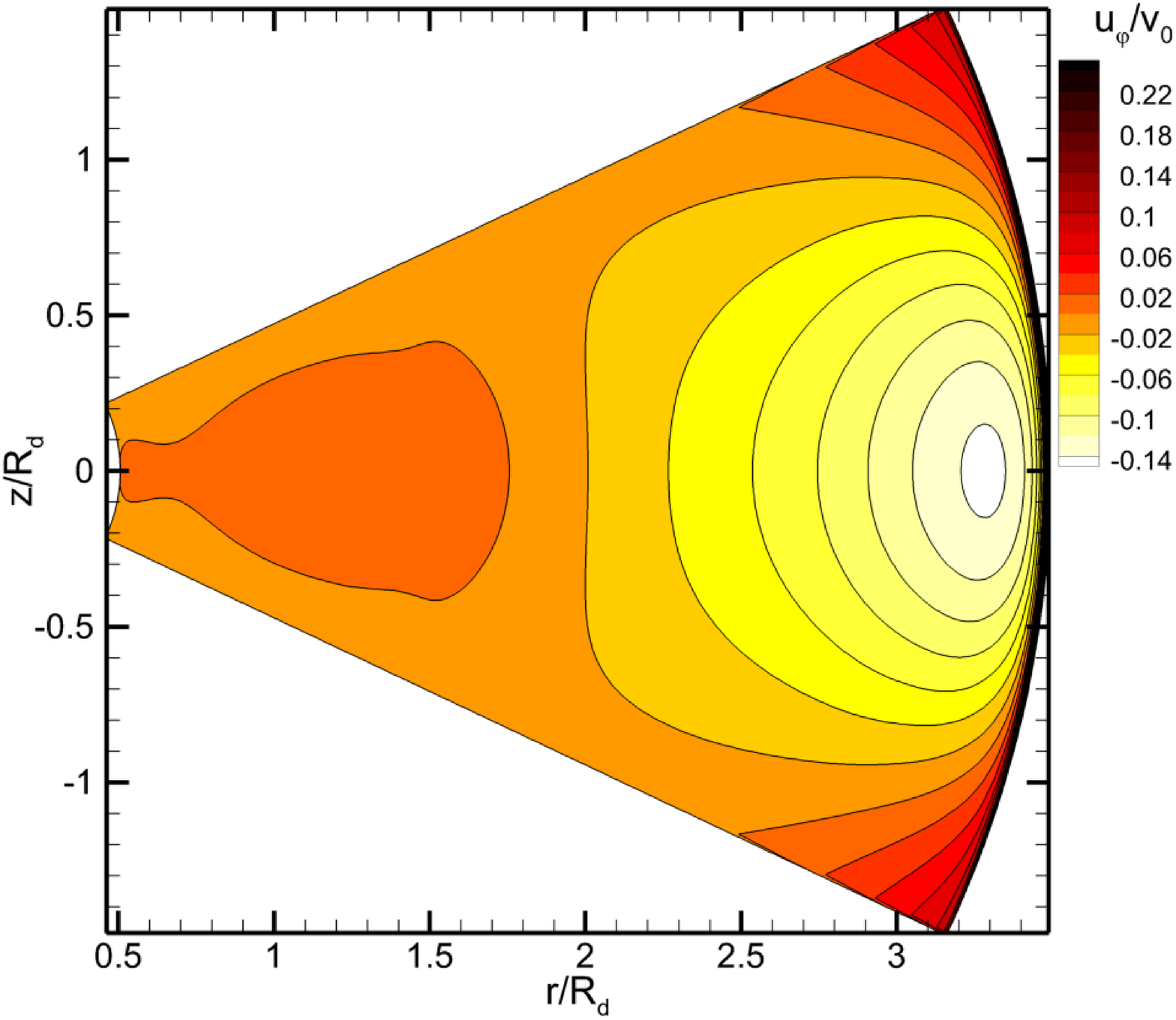}
\caption{Distribution of the velocity of the magnetic field lines $u_\varphi$ in Model 2 (temperature $T = 10^4$~K) at the initial time (left panel) and in the steady-state regime. The scaling factor is $v_0 = \sqrt{GM_\text{a}/R_\text{d}}$.}
\label{fig-u}
\end{figure}

Figure 10 shows numerical distributions of the scalar potential $\phi$ at the initial time (left panel) and at the onset of the stationary flow regime (right panel)
for Model 2 (temperature $T = 10^4$~K). The potential is given in dimensionless units. In this case, the scaling factor is $\phi_0 = \sqrt{4\pi\rho_\text{d}} GM_\text{a}/c$. As can be seen, the distribution of the potential changes significantly. While the maximum of the potential is initially located near the center of the computational domain, it is at the periphery in the stationary regime. This is due to the distribution of sources $\vec{B}_* \cdot \rotor\vec{v}$. At the initial time [see \eqref{eq-brotv2}], the source increases toward the center. In the stationary mode, the action of the electromagnetic force \eqref{eq-f2} leads to a deceleration of the rotation in the inner parts of the disk and in its corona. Therefore, the maximum of the source is shifted to the periphery.

The corresponding distributions of the velocity of the magnetic field lines $u_\varphi$ are shown in Fig. 11. The scaling factor is $v_0 = 
\sqrt{GM_\text{a}/R_\text{d}}$. At the initial time, velocity $u_\varphi$ increases toward the disk periphery. In the inner parts of the disk, $v_\varphi - u_\varphi > 0$, while $v_\varphi - u_\varphi < 0$ in the outer parts. This means that the electromagnetic force \eqref{eq-f2} tends to slow down the rotation of the inner parts of the disk and to unwind its outer parts. In the stationary regime, $v_\varphi - u_\varphi$ is positive everywhere. In other words, the action of the electromagnetic force leads to braking of the rotation. In this case, the contribution of the velocity of the magnetic field lines $u_\varphi$ to the magnitude of the electromagnetic force is approximately 25\%.

\end{document}